\documentclass[journal,11pt]{IEEEtran}

\usepackage{graphicx}
\usepackage{epsf}
\usepackage{amsmath, amsfonts}
\usepackage{latexsym}
\usepackage{multirow}
\usepackage{multicol}
\usepackage[table]{xcolor}
\usepackage[hidelinks,bookmarks=false]{hyperref}
\usepackage{bbm}
\usepackage{subcaption}

\begin{document}

\title{\fontsize{18}{8}\selectfont Human Electromagnetic Field Exposure in 5G at 28 GHz}

\author{
Seungmo Kim, \textit{Member, IEEE}, and Imtiaz Nasim

\vspace{-0.3 in}
}

\maketitle

\begin{abstract}
The fifth-generation wireless (5G) has already started showing its capability to achieve extremely fast data transfer, which makes itself considered to be a promising mobile technology. However, concerns have been raised on adverse health impacts that human users can experience in a 5G system by being exposed to electromagnetic fields (EMFs). This article investigates the human EMF exposure in a 5G system and compare to those measured in the previous-generation cellular systems. It suggests a minimum separation distance between a transmitter and a human user for keeping the EMF exposure below the safety regulation level, which provides consumers with a general understanding on the safe use of 5G communications.
\end{abstract}



\IEEEpeerreviewmaketitle

\section{Concern on Human EMF Exposure in 5G}\label{sec_intro}
As a means to fulfill the latest skyrocketing bandwidth demand, the fifth-generation wireless (5G) is expected to achieve far higher data rates compared to the previous-generation wireless systems. However, the 5G's requirement of a very high data rate entails an increase in signal power received at a user's end, which in turn results in an increase in the amount of electromagnetic energy imposed on the user \cite{5gappeal_sep17}-\cite{annal}. Not only that, this article identifies three technical features adopted in 5G, which can increase the human electromagnetic field (EMF) exposure `further.'

First, the 5G targets to operate at \textit{higher frequencies} (\textit{e.g.}, 28, 60, and 70 GHz \cite{jsac}-\cite{verboom20}) in addition to the existing lower-frequency bands for cellular communications. The advantages are (i) availability of wide bandwidths and (ii) possibility of integrating a larger number of antennas in small dimensions \cite{jsac}. At a higher frequency, however, the EMF `absorption' rate into human skin also rises.

Second, \textit{larger numbers of transmitters} will operate. In 5G, more base stations (BSs) will be deployed due to employment of small cells. As a direct consequence, BSs serve smaller geographic areas and thus are located closer to human users, which again results in a higher chance of a human user being exposed to EMF.

Third, \textit{directional beams} will be employed in 5G as a solution for faster attenuation of a signal power due to operation in high frequency bands \cite{jsac}. Notice that the main purpose of using such a multiple-antenna system is to increase the antenna gain. This higher concentration of electromagnetic energy results in a greater potential for an EMF to penetrate further into a human body.

\begin{table*}[t]
\caption{Parameters for case study in 5G, 4G, and 3.9G}
\centering
\begin{tabular}{|c|c|c|c|c}
\hline 
\textbf{Parameter} & \multicolumn{3}{|c|}{\textbf{Value}}\\ \hline \hline
&\multicolumn{1}{|c|}{\cellcolor{gray!10} 5G \cite{tr38901}} & \multicolumn{1}{|c|}{\cellcolor{gray!10}4G \cite{tr36873}} & \multicolumn{1}{|c|}{\cellcolor{gray!10}3.9G \cite{ts25996}}\\
\hline 
Carrier frequency & {28 GHz} & 2 GHz & 1.9 GHz\\
System layout & {Urban Macro (UMa)} & {Urban Macro (UMa)} & {Urban Macro (UMa)}\\
Inter-site distance (ISD) & 200 m & 500 m & 1 Km\\
Bandwidth & {850 MHz} & 20 MHz & 20 MHz\\ \hline
BS max antenna gain & {8 dBi per element} & 8 dBi per element &17 dBi\\
BS transmit power & 18 dBm per element & 44 dBm & 43 dBm\\
BS number of antennas (w/ separation of $\lambda/2$) & 256 and 64 & 4 & 4\\
BS antenna height & 25 m & 35 m & 32 m\\
BS noise figure & 5 dB & 5 dB & 5 dB\\ \hline
UE max antenna gain & 20 dBi & 1 dBi & 1 dBi\\
UE transmit power & 35 dBm & 23 dBm & 33 dBm\\
UE number of antennas (w/ separation of $\lambda/2$) & 16 & 4 & 1 (omni-directional)\\
UE antenna height & 1.5 m & 1.5 m & 1.5 m\\
UE noise figure & 9 dB & 9 dB & 9 dB\\ \hline
Cell sectorization & \multicolumn{3}{|c|}{3 sectors/site}\\
Deployment & \multicolumn{3}{|c|}{Outdoor 100\%} \\
Duplexing & \multicolumn{3}{|c|}{Time-division duplexing (TDD)} \\
Transmission scheme & \multicolumn{3}{|c|}{Single-user (SU)-MIMO} \\ \hline
\end{tabular}
\label{table_parameters}
\vspace{-0.1 in}
\end{table*}

\section{Current Understanding and Effort}\label{sec_related}
While substantial attention has been paid to technical advancements that the 5G will introduce, the potential impacts that the technology may pose on human health have not been discussed as closely and thoroughly.

\subsection{Health Effect}
`Heating' of skin is one representative impact on a human body caused by EMF exposure. The temperature for a skin outer surface normally ranges from 30 to 35$^{\circ}$C. The pain detection threshold temperature for human skin is approximately 43$^{\circ}$C \cite{wu15} and any temperature exceeding it can cause a long-term injury. Heating is considered as a significant impact since it can cause subsequent effects such as cell damage and protein induction \cite{pall18}. It is also known that high-frequency EMF affects the sweat glands (which may serve as helical antennas), peripheral nerves, the eyes and the testes, and may have indirect effects on many organs in the body \cite{book19}.

Recent studies showed health impacts of EMF in frequencies above 6 GHz. In a latest study \cite{kuster19}, EMF power transmitted to the body was analyzed as a function of angle of incidence and polarization, and its relevance to the current guidelines was discussed. Another study \cite{kuster18_bio} determined a maximum averaging area for power density (PD) that limits the maximum temperature increase to a given threshold. Also, considering `bursty' traffic patterns in modern wireless data communications, an analytical approach to `pulsed' heating was developed and applied to assess the peak-to-average temperature ratio as a function of the pulse fraction \cite{kuster18_phys}.

\subsection{Acknowledgement by Organizations}
The United States (US) Federal Communications Commission (FCC) \cite{fcc01} and the International Commission on Non-Ionizing Radiation Protection (ICNIRP) \cite{icnirp98} set guidelines on the maximum amount of EMF energy allowed on a human body. It is noteworthy that the FCC's guideline on specific absorption rate (SAR) is averaged over 1 gram (g) of tissue while that set by the ICNIRP is averaged over 10 g. It implies that the FCC's guideline is more conservative, while the ICNIRP allows for 2-3 times as much energy absorption.

Also, the US Food and Drug Administration (FDA) states that the current understanding on adverse impacts of EMF emissions on human health is insufficient to conclude whether exposure to the emissions is safe or not, and thus additional research is needed to address the current gaps in the literature on human health safety in use of wireless systems \cite{gao12}.

Meanwhile, the World Health Organization (WHO)'s International Agency for Research on Cancer (IARC) classifies EMF exposure as possibly carcinogenic \cite{WHO2011}.

\subsection{Measurements}
PD and SAR are the two most widely accepted metrics to measure the intensity and effects of EMF exposure \cite{em05}. However, selection of an appropriate metric evaluating the EMF exposure still remains controversial. The FCC suggests PD as a metric measuring the human exposure to EMF generated by devices operating at frequencies higher than 6 GHz \cite{fcc01}, whereas a recent study suggested that a guideline defined in PD is not efficient to determine the impacts on health issues especially when devices are operating in a very close proximity to the human body such as in an uplink \cite{wu15}.

However, PD cannot evaluate the effect of certain transmission characteristics (\textit{e.g.}, reflection) adequately. Thus, temperature elevation and SAR at a direct contact area are proposed as the appropriate metric for EMF exposure above 6 GHz \cite{temperature}. This article chooses SAR as a more adequate metric than the skin temperature, which is subject to be dispersed during propagation and affected by the external atmosphere (\textit{i.e.}, air temperature).

Every wireless device should pass compliance tests before going to the market. An international standard entitled IEC62232 \cite{iec62232} has been acting as a key reference in compliance tests for BSs and user equipments (UEs). It focuses on change of characteristics in adio frequency (RF) field with distance from a RF source. Relevant studies are also found. Exposure to RF EMF from a UE \cite{colombi18} and that from a BS \cite{colombi17} are studied.

\subsection{Reduction of Human EMF Exposure}
Albeit not many, schemes for EMF emission reduction in a wireless system have been studied \cite{sambo15}-\cite{nasim18}. Note that the human exposure can be reduced if a BS adopts a power control or adaptive beamforming technique \cite{baracca18}. Also, the exposure level can be reduced when multiple spectrum bands are combined for coordinated use. The reason is that with a higher carrier frequency, a wireless system should reduce the cell size, which leads to more severe threats to human health.

\subsection{Focus of This Article}
Four points on which this article puts particular focus are highlighted as follows. 

First, we discuss the human EMF exposure in the \textit{downlink} as well as the uplink. Most of the prior work studies the uplink only, while hardly paying attention to EMF emissions generated by BSs in a 5G network. Recall the aforementioned changes that the 5G adopts: (i) operation at higher carrier frequencies; (ii) reduction of cell size (which leads to increase in number of BSs); and (iii) concentration of higher EMF energy into an antenna beam. They all imply that in 5G, unlike the previous-generation wireless systems, the downlink can also be a threat to human health as well as the uplink.

Second, we suggest that \textit{both SAR and PD} should be used to display human EMF exposure for a wireless system. The reason is that SAR captures an amount of EMF energy that is actually `absorbed' into human tissues, while PD is an efficient metric only to present the EMF energy being introduced to a human user.

Third, we present an \textit{explicit comparison} of human EMF exposure in 5G to those in the currently deployed wireless standards. For 5G, we adopt the system model defined in the Third-Generation Partnership Project (3GPP) 5G New Radio (NR) \cite{tr38901}. Meanwhile, currently operating technologies are represented by 4G \cite{tr36873} and 3.9G \cite{ts25996}. Notice that 4G represents the 3GPP's Long Term Evolution (LTE)-Advanced, and 3.9G is the last release by the 3GPP before 4G was deployed (from which the name ``3.9G'' is originated).

Fourth, we consider the maximum possible exposure that a human user can experience. In other words, no technique for mitigation of received power is considered in this paper's system model. It is for advising the consumers with most conservative perspectives in using 5G wireless.

\begin{figure}
\begin{minipage}[t]{\linewidth}
\centering
\includegraphics[width = .9\linewidth]{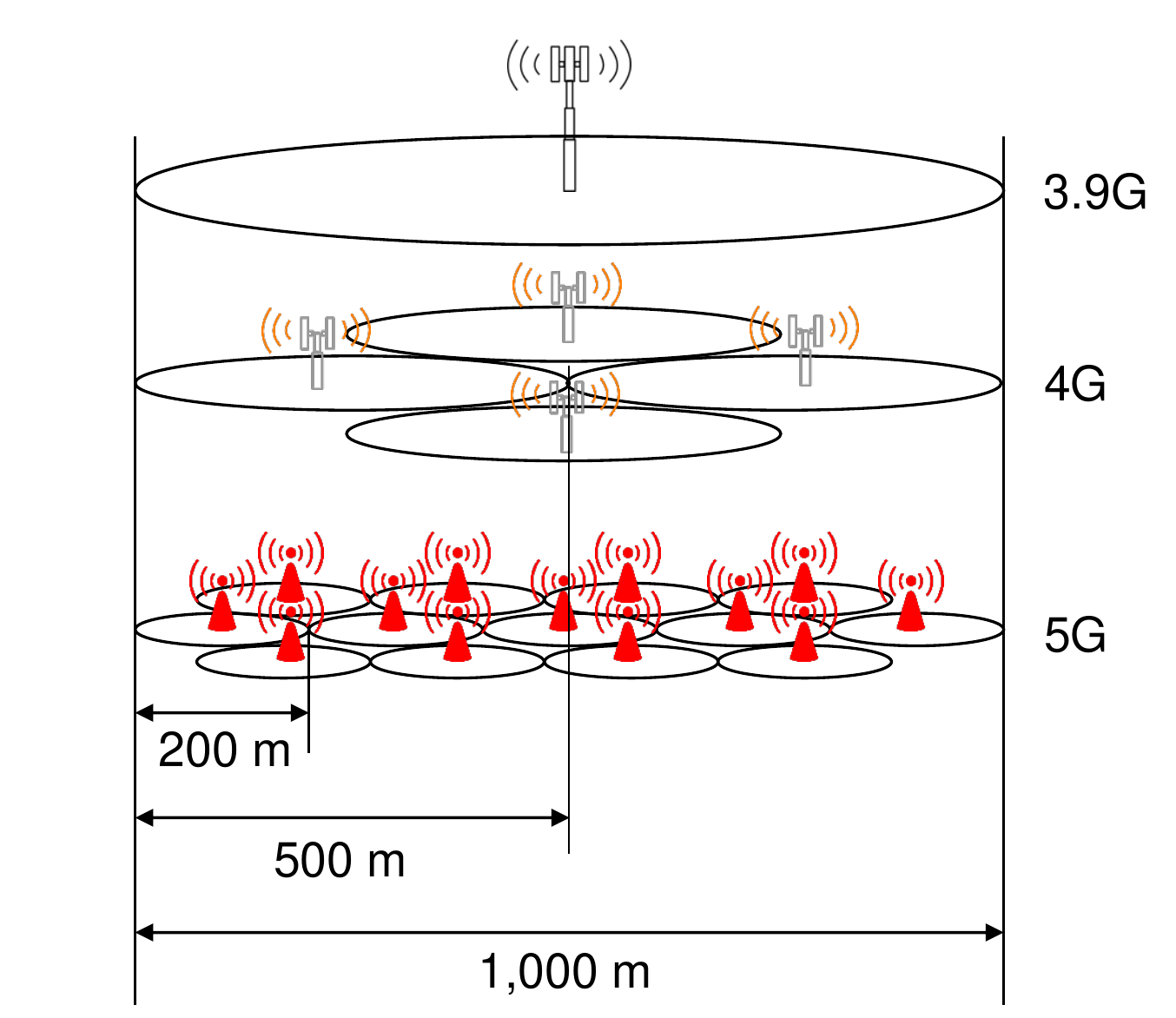}
\caption{Comparison of cell size}
\label{fig_comparison_size}
\vspace{0.2 in}
\end{minipage}
\begin{minipage}[t]{\linewidth}
\centering
\includegraphics[width = .9\linewidth]{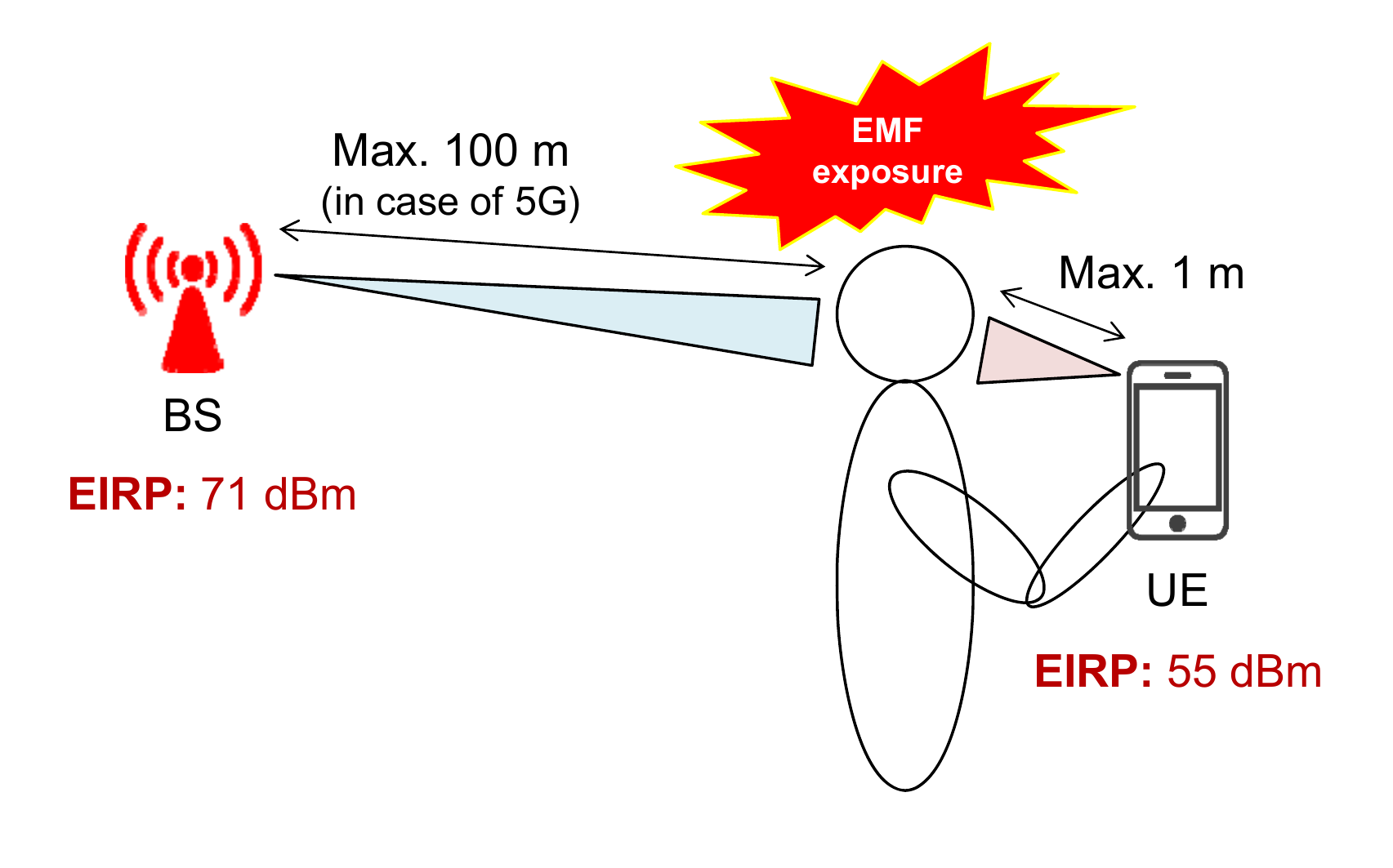}
\caption{Comparison of uplink and downlink in 5G}
\label{fig_comparison_uplink}
\end{minipage}
\end{figure}

\begin{figure*}
\centering
\begin{subfigure}{.45\textwidth}
\centering
\includegraphics[width = \textwidth]{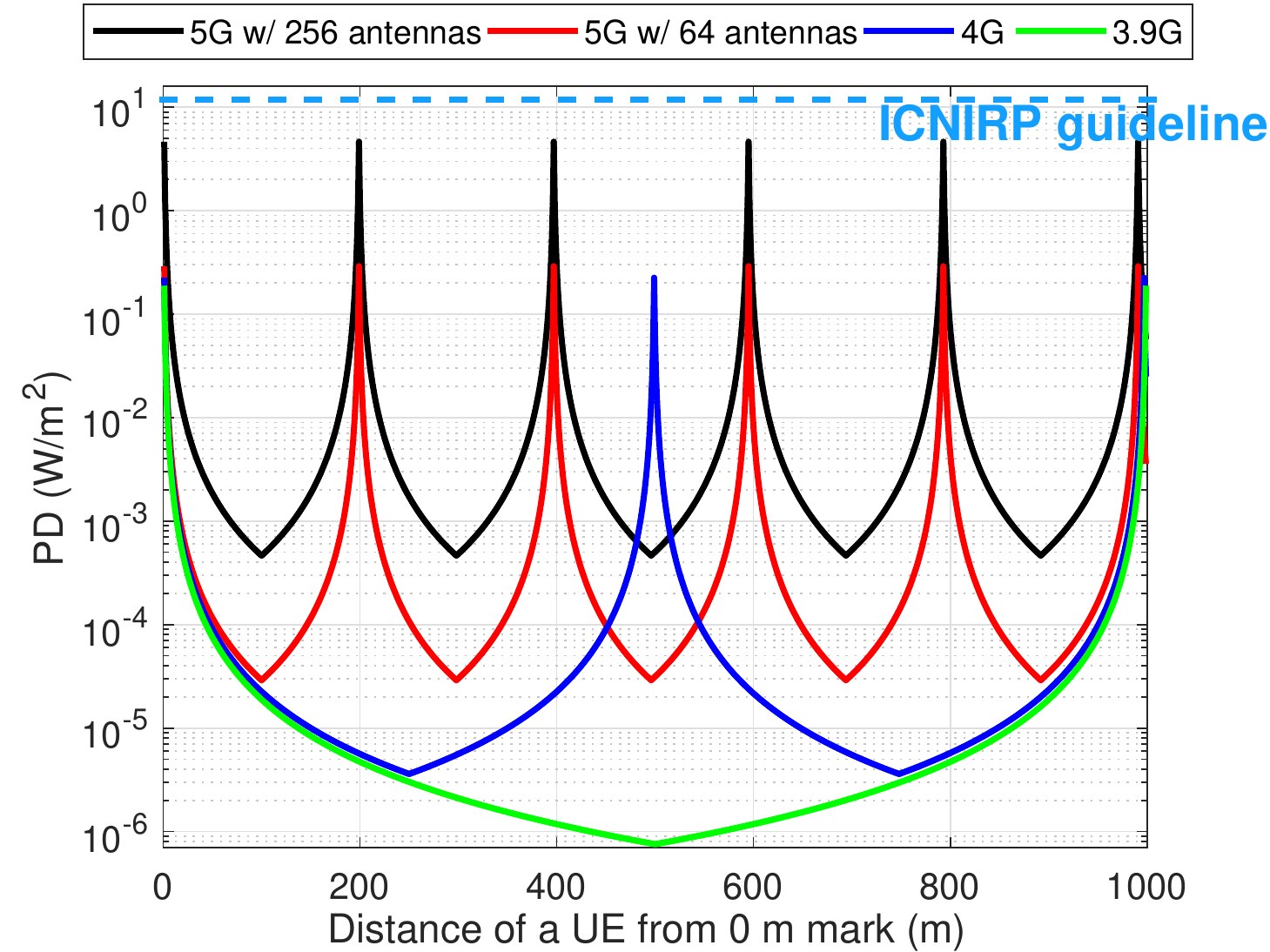}
\caption{Comparison of time-averaged PD in downlink}
\label{fig_pd_dl}
\vspace{0.2 in}
\end{subfigure}
\begin{subfigure}{.45\textwidth}
\centering
\includegraphics[width= \textwidth]{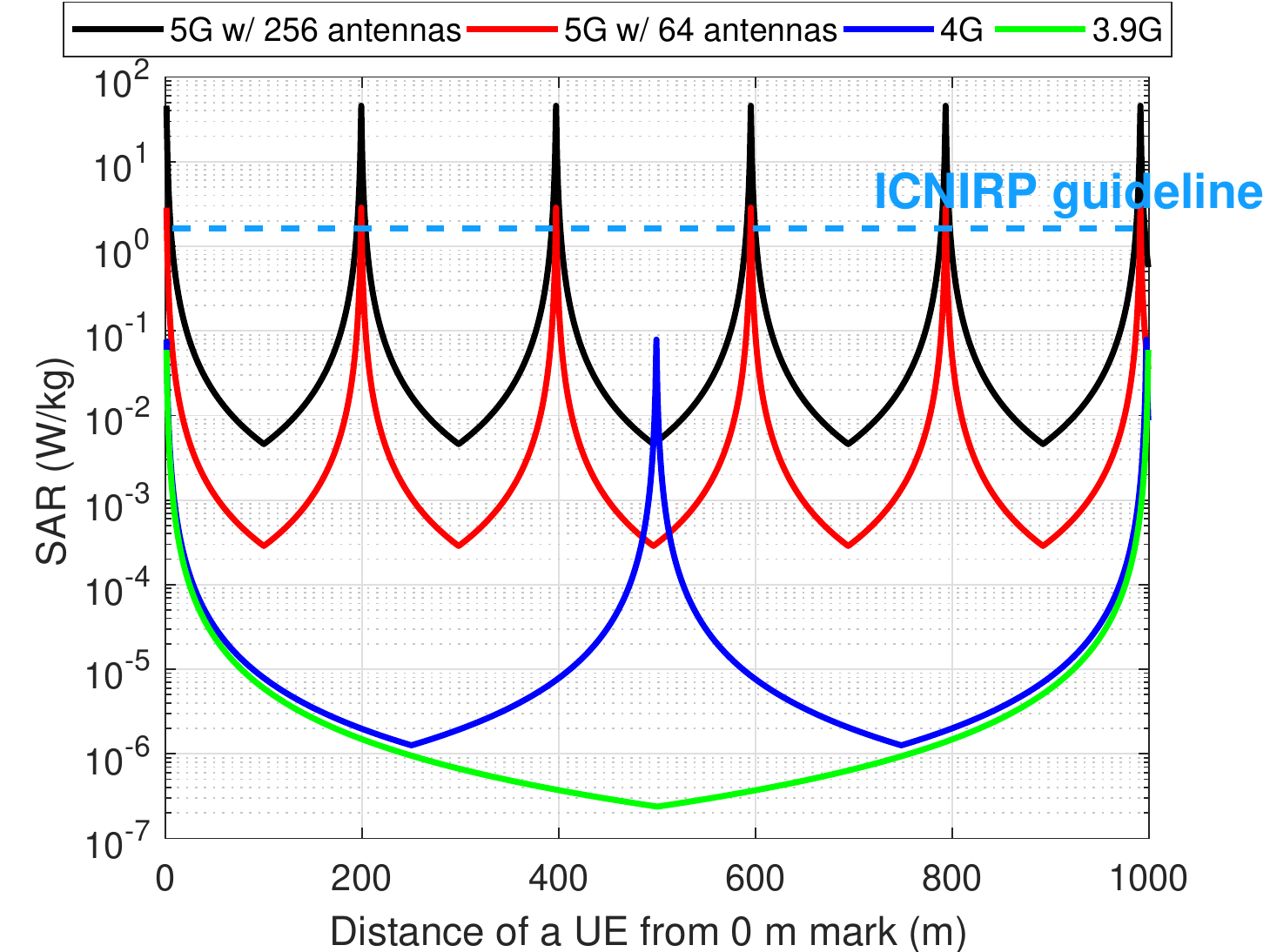}
\caption{Comparison of time-averaged SAR in downlink}
\label{fig_sar_dl}
\vspace{0.2 in}
\end{subfigure}
\begin{subfigure}{.45\textwidth}
\centering
\includegraphics[width=\textwidth]{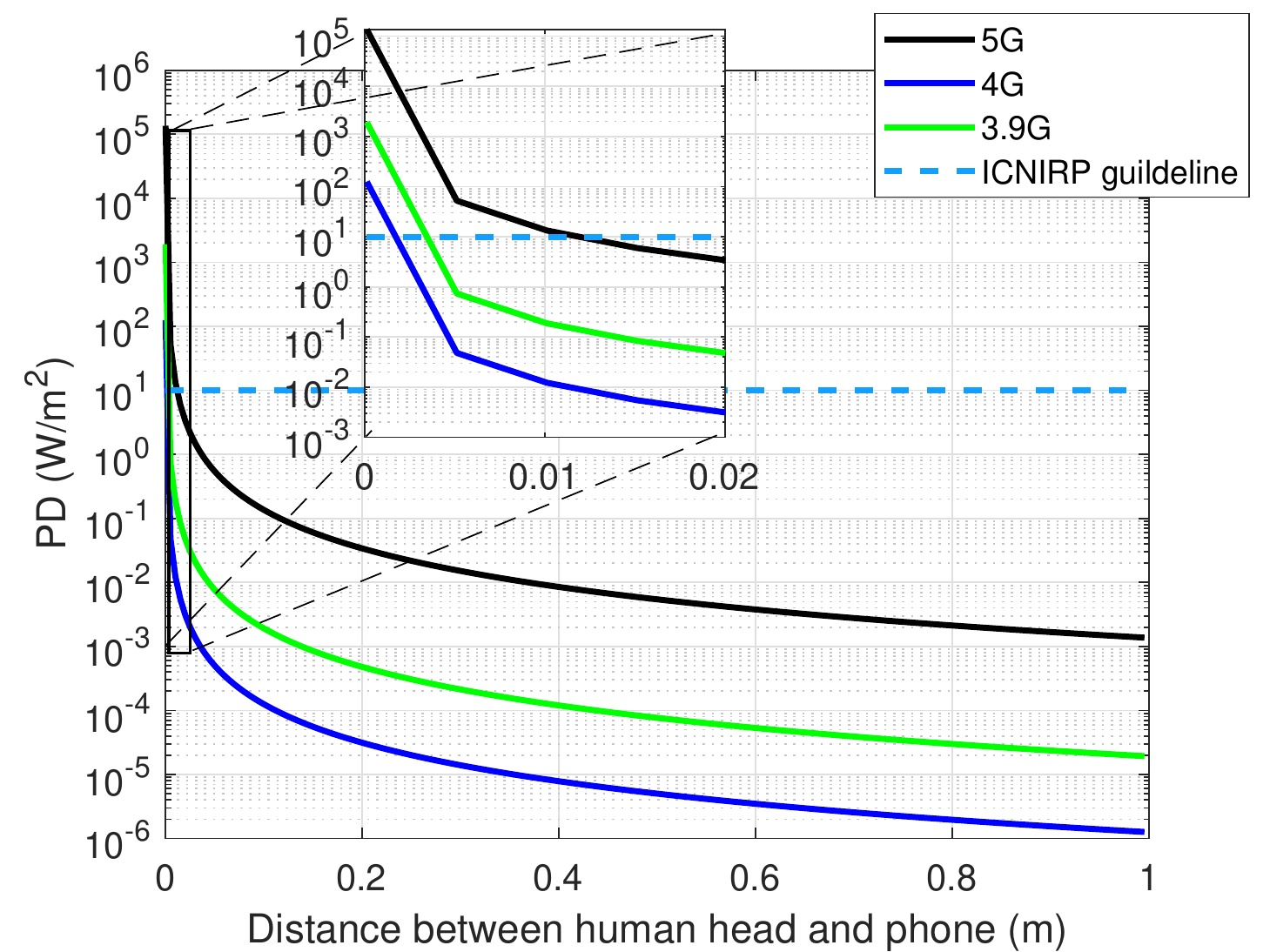}
\caption{Comparison of time-averaged PD in uplink}
\label{fig_pd_ul}
\end{subfigure}
\begin{subfigure}{.45\textwidth}
\centering
\includegraphics[width=\textwidth]{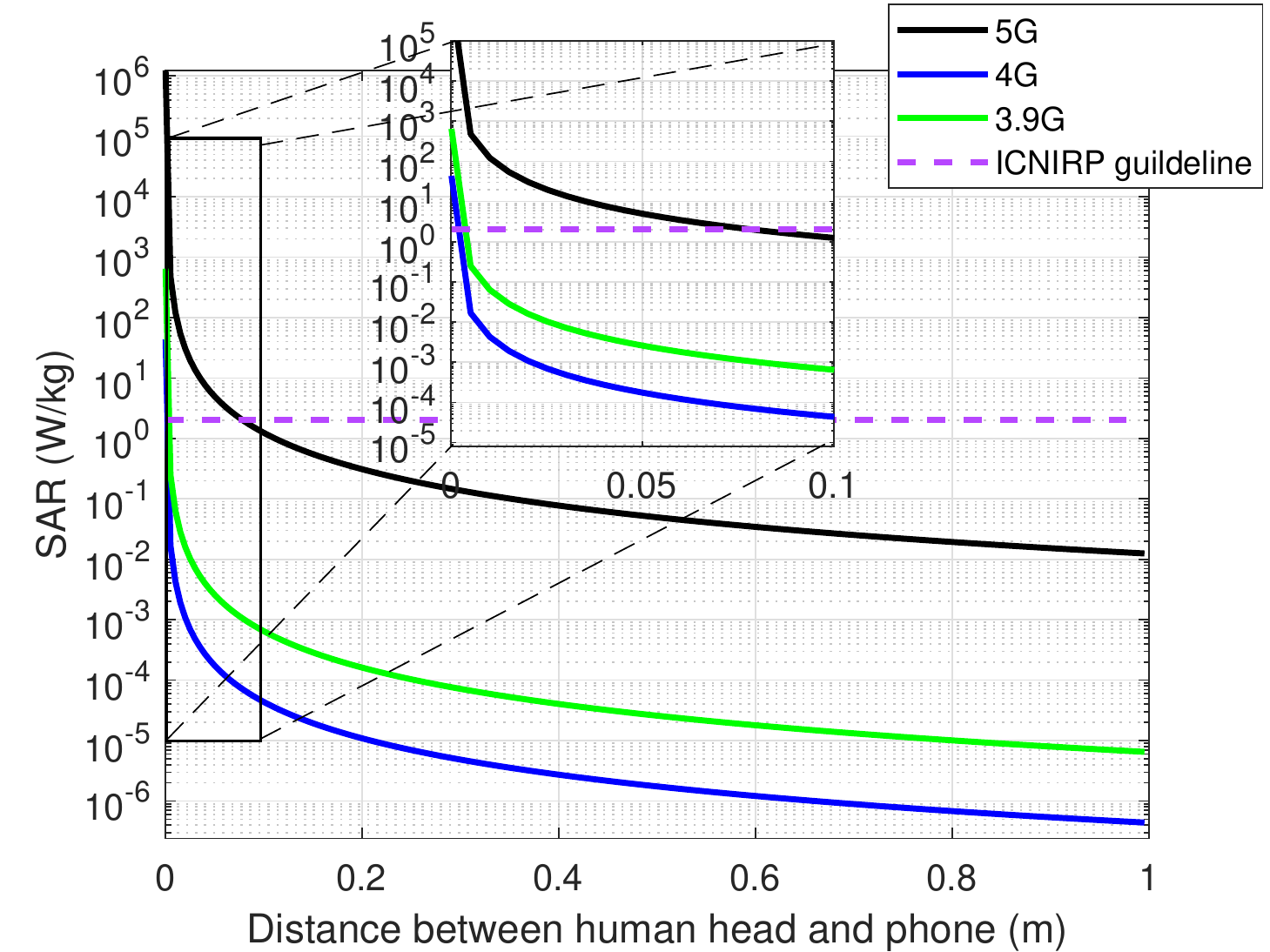}
\caption{Comparison of time-averaged SAR in uplink}
\label{fig_sar_ul}
\end{subfigure}
\vspace{0.1 in}
\caption{Comparison of human exposure levels}
\vspace{-0.1 in}
\end{figure*}

\section{Case Study Models}\label{sec_model}
In order to understand how much EMF energy is imposed to a human user in a 5G wireless system, this article suggests two `comparative' case study models: (i) among different wireless systems--\textit{i.e.}, 5G, 4G, and 3.9G, and (ii) between downlink and uplink.

\subsection{5G vs. 4G vs. 3.9G}
Commonly for all of the three systems, we assume a fully loaded network in order to understand the `worst case' of EMF exposure. As mentioned earlier, none of the three systems is supposed to adopt any `adaptive' techniques--\textit{viz.}, power control and adaptive beamforming. That is, there is no particular method applied to reduce the amount of EMF energy being imposed to a user at a certain time instant. The rationale is to provide the `most conservative' suggestion on consumer safety, leaving room for a safety margin \cite{colombi18}.

Fig. \ref{fig_comparison_size} depicts difference in cell size among the three wireless standards. As mentioned in Section \ref{sec_intro}, a 5G system adopts the smallest cell diameter (\textit{i.e.}, 200 m) among the three systems, pursing to form a small-cell network. This difference in cell size is a significant factor differentiating the level of human EMF exposure among 5G, 4G, and 3.9G, as shall be discussed in Section \ref{sec_results}.

\subsection{Downlink vs. Uplink}
Another case study is defined as comparison between uplink and downlink in a 5G system. Fig. \ref{fig_comparison_uplink} illustrates the geometric difference between the two directions of communication. In this case study, the user's head is placed between the BS and the handheld device, which represents a case where the impact of human EMF exposure is highlighted.

There is a key similarity between the downlink and the uplink: both adopt beamforming \cite{tr38901}. Accordingly, they both adopt directional antennas, which results in concentration of electromagnetic energy higher than 0 dB in an antenna beam.

The differences between downlink and uplink are as follows. First, in uplink, the equivalent isotropically radiated power (EIRP) that a transmitter generates is lower than in a downlink. The reason is two-fold: (i) an uplink requires a lower data rate than a downlink; and (ii) a UE, as a transmitter, is less capable of accommodating as many antennas as a BS can. Second, a signal propagates shorter in an uplink than in a downlink. The inter-site distance (ISD) for a 5G cell is 200 m, as indicated in Table \ref{table_parameters}, which yields a cell radius to be 100 m. As a consequence, in downlink, the maximum distance that a user can be separated from a BS is 100 m. In contrast, in an uplink scenario, the maximum separation distance from the human user and the transmitter (a handheld device as being in an uplink) is supposed to be 1 m. When a handheld device is held in a user's hands, one can consider a number of representative scenarios such as directly contacting at an ear, moderate separation for texting or web surfing, and further separation with use of an ear bud. The `maximum 1 m' in Fig. \ref{fig_comparison_uplink} came from assumption of the last scenario that yields the maximum distance between the handheld device and the user's head. Third, an antenna beam in an uplink is less strong and sharp than in a downlink. This is associated with the aforementioned geometric difference: a downlink beam is designed stronger and sharper for overcoming larger attenuation through a longer propagation.

\section{Numerical Results and Discussions}\label{sec_results}
Now we evaluate human EMF exposure for the three wireless systems (\textit{i.e.}, 5G, 4G, and 3.9G) via Monte Carlo simulations in the case studies defined in Section \ref{sec_model}. Specifically, to consider variation of a mobile user's relative location in a cell, both of PD and SAR are `averaged' in 10,000 experiments, each of which generates 10 UEs per sector. Also, a cell is assumed fully loaded; the calculation considers a time length that is enough for all the 10 UEs that are served based on TDD.

\subsection{5G vs. 4G vs. 3.9G}
In simulation of the first case study, the Urban Macro (UMa) system layout is assumed, which is commonly defined in all of the three wireless standards that this article refers to, as already shown in Table \ref{table_parameters}. See the following specifications for technical details: 5G \cite{tr38901}, 4G \cite{tr36873}, and 3.9G \cite{ts25996}.

In Figs. \ref{fig_pd_dl} and \ref{fig_sar_dl}, there is a BS located at `0 m' mark for all 5G, 4G, and 3.9G systems. Now, a mobile user is moved from the 0 m mark to the 1,000 m mark. Since each of 5G, 4G, and 3.9 system adopts a different cell radius (also known as ISD), the downlink signal that the UE receives gets bounced up as it passes another BS standing at different distance marks. This experiment setting is to highlight the impact of adopting smaller cells in 5G. Comparing 5G to 4G in both of Figs. \ref{fig_pd_dl} and \ref{fig_sar_dl}, despite faster attenuation than 4G due to operation at a higher frequency, PD and SAR in 5G are kept elevated more frequently, as the UE meets the next BS in a shorter distance. That is, in a 5G network, a consumer is likely to be exposed to more consistently high EMF energy. Nevertheless, it is easier to apply a `compliance distance' \cite{iec62232} in a downlink than in an uplink. Thus, this article suggests (i) an overhaul of the compliance distances defined in different standards and (ii) the consumers' discretion on being close to a BS.

Compare black and blue curves in Figs. \ref{fig_pd_dl} and \ref{fig_sar_dl}. It is evident that the difference between 5G (with 256 antennas) and 4G is larger in SAR than in PD. This is explained by formal expression of $\text{SAR}\left(d,\phi\right) = 2\text{PD}\left(d, \phi\right) \left(1 - \mathcal{R}^2\right)/\left(\delta \rho\right)$ where $\mathcal{R}$ is the reflection coefficient \cite{wu15}; $\rho$ is the tissue mass density (1 $\text{g}/\text{cm}^3$ is used); and $\delta$ is the skin penetration depth (10$^{\text{-3}}$ m is used) \cite{wu15}. Recall that 5G and 4G operate at 28 GHz and 2 GHz, respectively. The SAR is inversely proportional to the penetration depth, and hence a shallower penetration occurring in 5G yields a higher absorption.

\begin{figure}[t]
\centering
\includegraphics[width = \linewidth]{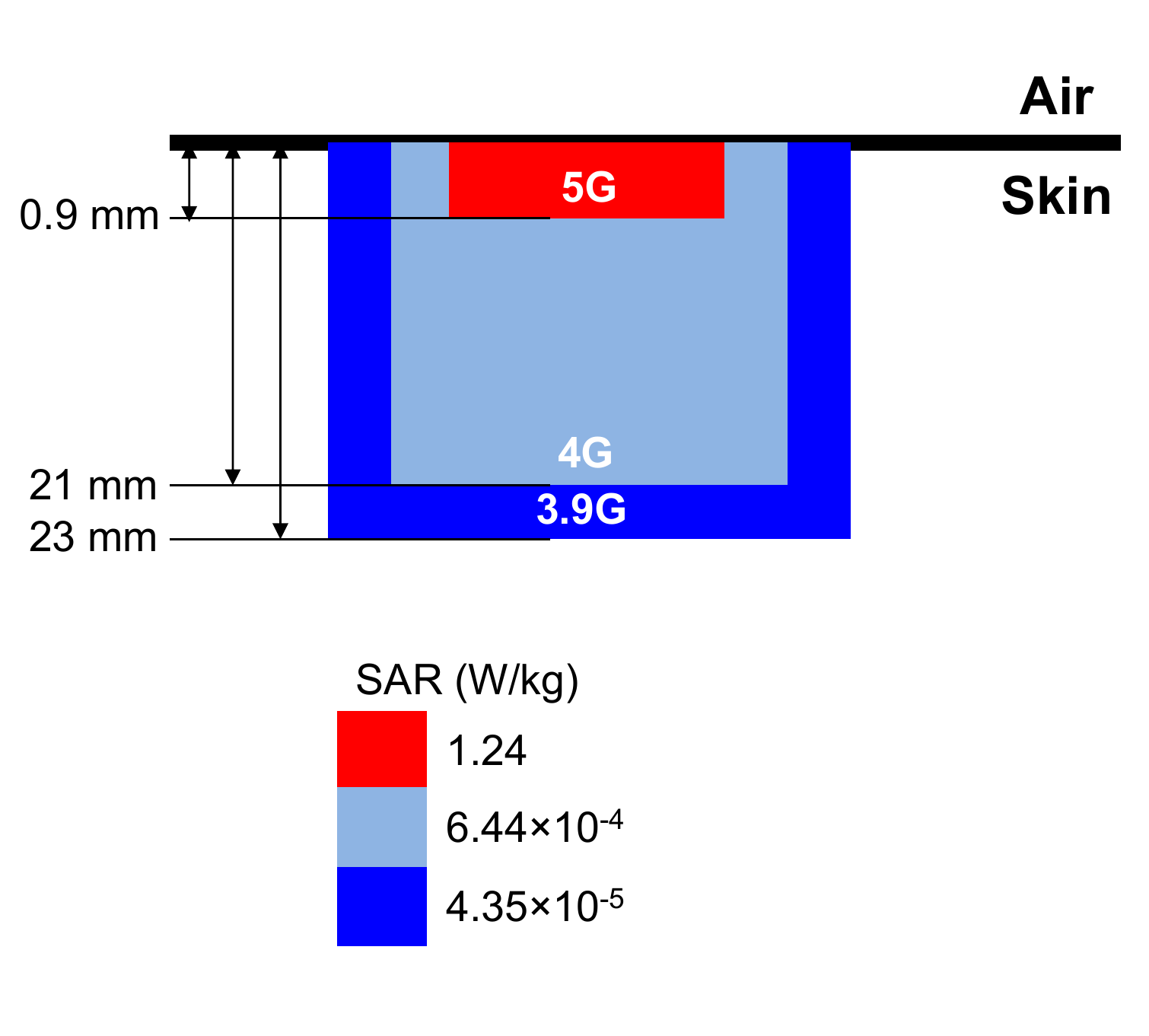}
\caption{Comparison of EMF penetration depth into human skin and SAR}
\label{fig_depth}
\end{figure}

Fig. \ref{fig_depth} compares the depth that an EMF penetrates into human skin, among the three wireless systems of 5G, 4G, and 3.9G. Note that the level of SAR varies according to a number of disparate variables--\textit{i.e.}, type of material, frequency, etc. The example shown in Fig. \ref{fig_depth} presents a measurement of SAR being introduced on human skin at the distance of 10 cm from a transmitter in an uplink. It clearly shows a scenario where the current belief is not valid; the fact that a high-frequency EMF cannot penetrate deep into human skin does not mean that it is not dangerous. Specifically, although the penetration is limited only at the skin surface, the SAR (illustrated as a heat map in Fig. \ref{fig_depth}) can be higher within the concentrated area, which can cause subsequent health problems such as skin heating.

\subsection{Downlink vs. Uplink}
Figs. \ref{fig_pd_ul} and \ref{fig_sar_ul} compare PD and SAR in uplink to the ICNIRP guidelines set at 10 W/m$^2$ and 2 W/kg, respectively. PD and SAR are remarkably higher in uplink than downlink, shown via comparison of the results for uplink to those for downlink shown in Figs. \ref{fig_pd_dl} and \ref{fig_sar_dl}. It is attributed to smaller separation distance between a transmitter and a human body. Imagine one talking on a voice call; it is a `direct' physical contact of the phone and the head! It has been found that the \textit{distance} between the EMF emitting device and the human body is a factor predominating the amount of the EMF energy absorption into the body \cite{milcom19}\cite{arxiv19}.

Also, it is significant to notice that no regulation exists at 28 GHz where this article investigates for 5G. As such, we refer to the ICNIRP's guideline that is set to be 2 W/kg by the \cite{icnirp98} at a frequency `below 10 GHz.' In Fig. \ref{fig_sar_ul}, it provides a ``inferred'' understanding on SAR in an uplink. The zoom-in look shown in Fig. \ref{fig_sar_ul} suggests that in 5G, use of a handheld device within the distance of 8 cm causes an EMF absorption exceeding 2 W/kg, which would have been prohibited if the carrier frequency was lower than 10 GHz. This implies the gravity of human EMF exposure in an uplink of 5G.

\section{Conclusions}
This article has discussed human EMF exposure in 5G operating at 28 GHz, while most of the prior work focuses only on the technological benefits that the technology brings. Considering the significance of wireless technologies in our daily life, the potential danger of using them should also be emphasized for sustainable advancement of the technologies. In this article, the first case study has demonstrated how much EMF exposure is caused in a 5G system compared to 4G and 3.9G. Then the latter case study has suggested an adequate separation distance from a transmitter, in order to keep a human user from being exposed to EMF below a regulatory guideline. This article is expected to ignite continued interest in overarching research on the design of future wireless systems that achieve high performance while keeping consumer safety guaranteed.

However, considering the gravity of this issue, we suggest several directions to be achieved in our future research:
\begin{itemize}
\item \textit{Human EMF exposure mitigation strategy:} We are particulary interested in exploiting the technical features in future wireless systems--\textit{i.e.}, a larger number of BSs within a unit area. Such a paradigm change will enable a holistic, network-based approach to mitigate the EMF exposure as an optimization problem with a set of constraints representing the PD, SAR, and skin temperature elevation.
\item \textit{Further studies regarding exact human health impacts caused by EMF expousre:} The particular focus will be put on (i) skin dielectric effect with respect to frequency and (ii) the effect of radiation when the body is covered with clothing or garment materials.
\end{itemize}

\section{About the Authors}
S. Kim (seungmokim@georgiasouthern.edu) is an assistant professor in the Department of Electrical and Computer Engineering at Georgia Southern University, and the director of the New-Era Wireless (NEW) Laboratory.

I. Nasim (in00206@georgiasouthern.edu) earned his M.S. in electrical engineering at Georgia Southern University in 2019.



\begin{thebibliography}{99}
\setlength{\parskip}{0.000000001em}
\bibitem{5gappeal_sep17} ``Scientists warn of potential serious health effects of 5G,'' [Online]. Available: \url{https://ehtrust.org/wp-content/uploads/Scientist-5G-appeal-2017.pdf}

\bibitem{southeast} I. Nasim and S. Kim, ``Adverse impacts of 5G downlinks on human body,'' in \textit{Proc. IEEE SoutheastCon 2019}.

\bibitem{annal} I. Nasim and S. Kim, ``Mitigation of human EMF exposure in downlink of 5G,'' \textit{Springer Annals of Telecommun.}, vol. 74, iss. 1-2, Feb. 2019.

\bibitem{jsac} S. Kim, E. Visotsky, P. Moorut, K. Bechta, A. Ghosh, and C. Dietrich, ``Coexistence of 5G with the Incumbents in the 28 and 70 GHz Bands,'' \textit{IEEE J. Sel. Areas Commun.}, vol. 35, no. 6, Jun. 2017.

\bibitem{hst19} M. Kabir and S. Kim, ``5G or Wi-Fi for HA/DR in the 60 GHz Band?,'' in \textit{Proc. IEEE Int. Symp. Technol. Homeland Security 2019}.

\bibitem{verboom20} J. Verboom and S. Kim, ``Stochastic analysis on downlink performance of coexistence between WiGig and NR-U in 60 GHz band,'' \textit{arXiv:2003.01570}, Mar. 2020.

\bibitem{pall18} M. Pall, ``Wi-Fi is an important threat to human health,'' \textit{Elsevier Environmental Research}, vol. 164, Mar. 2018.

\bibitem{book19} M. Markov, \textit{Mobile communications and public health}, CRC Press, 2019.

\bibitem{kuster19} T. Samaras and N. Kuster, ``Theoretical evaluation of the power transmitted to the body as a function of angle of incidence and polarization at frequencies $>$6 GHz and its relevance for standardization,'' \textit{Bioelectromagn.}, vol. 40, no. 2, Feb. 2019.

\bibitem{kuster18_bio} E. Neufeld, E. Carrasco, M. Murbach, Q. Balzano, A. Christ, and N. Kuster, ``Theoretical and numerical assessment of maximally allowable power-density averaging area for conservative electromagnetic exposure assessment above 6 GHz,'' \textit{Bioelectromagn.}, vol. 39, no. 8, Dec. 2018.

\bibitem{kuster18_phys} E. Neufeld and N. Kuster, ``Systematic derivation of safety limits for time-varying 5G radiofrequency exposure based on analytical models and thermal dose,'' \textit{Health Phys.}, Sep. 2018.

\bibitem{fcc01} FCC, ``Evaluating compliance with FCC guidelines for human exposure to radiofrequency electromagnetic fields,'' \textit{Supplement C Edition 01-01 to OET Bulletin 65 Edition 97-01}, Jun. 2001.

\bibitem{icnirp98} ICNIRP, ``ICNIRP guidelines: for limiting exposure to time-varying electric, magnetic and electromagnetic fields (100 kHz to 300 GHz),'' Jul. 2018. [Online]. Available: \url{https://www.icnirp.org/cms/upload/consultation_upload/ICNIRP_RF_Guidelines_PCD_2018_07_11.pdf}

\bibitem{gao12} United States Government Accountability Office, ``Telecommunications: exposure and testing requirements for mobile phones should be reassessed,'' \textit{GAO-12-771}, Aug. 2012.

\bibitem{WHO2011} The U.S. Food and Drug Administration (FDA), \textit{Current research result on cell phones}, updated Mar. 2018.

\bibitem{wu15} T. Wu, T. Rappaport, and C. Collins, ``Safe for generations to come: considerations of safety for millimeter waves in wireless communications,'' \textit{IEEE Microw. Mag.}, vol. 16, no. 2, 2015.

\bibitem{em05} \textit{Draft standard for safety levels with respect to human exposure to the radio frequency electromagnetic fields, 0 Hz to 300 GHz}, IEEE Standard PC95.1/D3.5, Oct. 2018.

\bibitem{temperature} M. Ziskin, S. Alekseev, K. Foster, and Q. Balzano ``Tissue models for RF exposure evaluation at frequencies above 6 GHz,'' \textit{Bioelectromagn.}, vol. 39, no. 3, Apr. 2018.


\bibitem{iec62232} IEC, \textit{Determination of RF field strength, power density and SAR in the vicinity of radiocommunication base stations for the purpose of evaluating human exposure}, IEC 62232 ED2, Aug. 2017.

\bibitem{colombi18} D. Colombi, B. Thors, C. Tornevik, and Q. Balzano, ``RF energy absorption by biological tissues in close proximity to millimeter-wave 5G wireless equipment,'' \textit{IEEE Access}, vol. 6, Aug. 2018.

\bibitem{colombi17} B. Thors, A. Furuskar, D. Colombi, and C. Tornevik, ``Time-averaged realistic maximum power levels for the assessment of radio rrequency exposure for 5G radio base stations using massive MIMO,''' \textit{IEEE Access}, vol. 5, Oct. 2017.

\bibitem{sambo15} Y. A. Sambo, F. Heliot, and M. Imran, ``Electromagnetic emission-aware scheduling for the uplink of coordinated OFDM wireless systems,'' in \textit{Proc. IEEE Online Conf. Green Commun. (OnlineGreenComm)}, 2015.

\bibitem{love16} M. Castellanos, D. Love, and B. Hochwald, ``Hybrid precoding for millimeter wave systems with a constraint on user electromagnetic radiation exposure,'' in \textit{Proc. Asilomar Conf. Signals, Syst. and Comput.}, Nov. 2016.

\bibitem{nasim18} I. Nasim and S. Kim, ``Mitigation of human RF exposure in 5G downlink,'' \textit{arXiv: 1807,09094}, Jul. 2018.

\bibitem{baracca18} P. Baracca, A. Weber, T. Wild, and C. Grangeat, ``A statistical approach for RF exposure compliance boundary assessment in massive MIMO systems,'' in \textit{Int. ITG Workshop Smart Ant.}, Mar. 2018.

\bibitem{tr38901} 3GPP TR 38.901, ``Channel model for frequencies from 0.5 to 100 GHz (Release 14),'' v14.3.0, Dec. 2017.

\bibitem{tr36873} 3GPP TR 36.873, ``Study on 3D channel model for LTE (Release 12),'' v12.0.0, Sep. 2014.

\bibitem{ts25996} 3GPP TR 25.996, ``Spatial channel model for multiple input multi output (MIMO) simulations (Release 9),'' v9.0.0, Dec. 2009.

\bibitem{milcom19} I. Nasim and S. Kim, ``Human EMF exposure in wearable networks for internet of battlefield things,'' in \textit{Proc. IEEE Military Commun. Conf. (MILCOM) 2019}.

\bibitem{arxiv19} S. Kim, Y. Sharif, and I. Nasim, ``Human electromagnetic field exposure in wearable communications: a review,'' \textit{arXiv:1912.05282}, Dec. 2019. 

\end{thebibliography}
\end{document}